\documentclass[journal=jacsat,manuscript=article]{achemso}

\usepackage[version=3]{mhchem} 

\author{J. Binder}
\email{johannes.binder@fuw.edu.pl}
\author{A. K. Dabrowska}
\author{M. Tokarczyk}
\author{K. Ludwiczak}
\author{R. Bozek}

\author{G. Kowalski}
\author{R. Stepniewski}
\author{A. Wysmolek}

\affiliation{Faculty of Physics, University of Warsaw, ul. Pasteura 5, 02-093 Warsaw, Poland}

\title{Epitaxial hexagonal boron nitride for hydrogen generation by radiolysis of interfacial water}

\keywords{hexagonal boron nitride, bubble, two-dimensional materials, hydrogen production, hydrogen storage, Raman spectroscopy, hydrogen barrier, radiolysis}

\begin{document}



\begin{abstract}
  Hydrogen is an important building block in global strategies towards a future green energy system \cite{Renssen2020,Griffiths2021}. To make this transition possible, intense scientific efforts are needed, also in the field of materials science. Two-dimensional crystals, such as hexagonal boron nitride (hBN)\cite{Cassabois2016}, are very promising in this regard, as it was demonstrated that micrometer-sized exfoliated flakes are excellent barriers to molecular hydrogen\cite{He2019, Blundo2022}. However, it remains an open question whether large-area layers fabricated by industrially relevant methods preserve such promising properties. In this work we show that electron beam-induced splitting of water creates hBN bubbles that effectively store molecular hydrogen for weeks and under extreme mechanical deformation. We demonstrate that epitaxial hBN allows direct visualization and monitoring of the process of hydrogen generation by radiolysis of interfacial water. Our findings show that hBN is not only a potential candidate for hydrogen storage, but also holds promise for the development of unconventional hydrogen production schemes.

\end{abstract}

Although hydrogen is the most abundant element in the solar system, its production and storage constitute a major scientific challenge that hinders its widespread application, for example, as a fuel or feedstock. It has been experimentally shown that hBN holds great prospects for hydrogen applications because it is an excellent proton conductor \cite{Hu2014,Yoon2018}, a molecular hydrogen barrier\cite{He2019, Blundo2022} and an electrical insulator \cite{Britnell2012} that can withstand high temperatures and harsh environments\cite{Paine1990,Roy2021}. Moreover, hBN shows exceptional mechanical properties \cite{Falin2017}, which are currently being explored in fields like flexible electronics\cite{Moon2022}. The main challenge is to preserve the above properties while scaling from submillimeter-sized mechanically exfoliated samples to thin films with a large area \cite{KidambiPiran2022}. 

\begin{figure}[h!]
	\centering
		\includegraphics[width=0.95\textwidth]{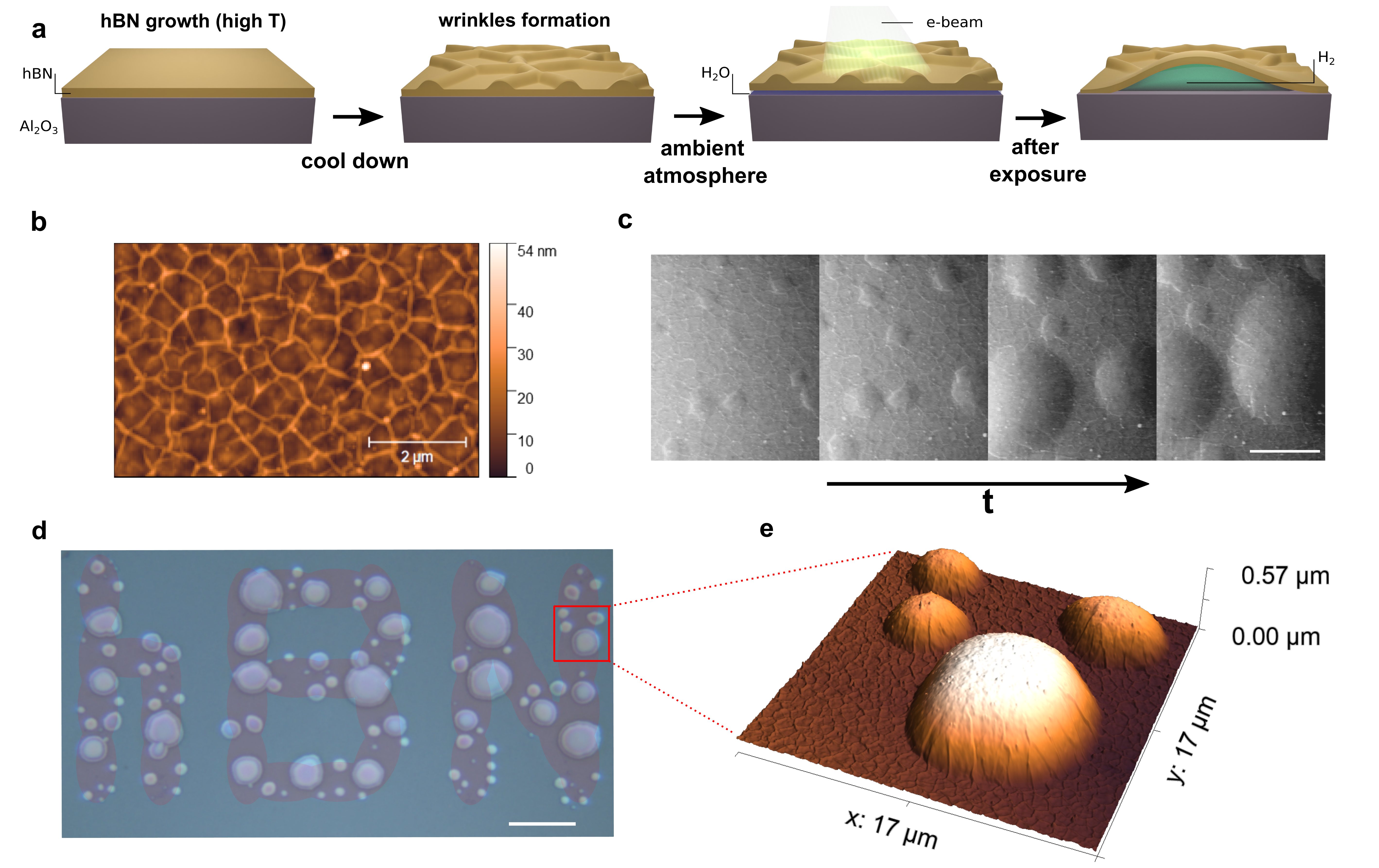}
	\caption{Bubble formation mechanism. (a) Schematic illustration of the bubble formation. hBN is grown by MOVPE at temperatures above 1000 $^{\circ}$C. After the growth the sample is cooled down to room temperature, which leads to the formation of hBN wrinkles. The sample is removed from the reactor and exposed to ambient conditions. Electron beam exposure in an SEM leads to bubble formation. (b) shows an AFM image of a typical wrinkle pattern (c) shows the evolution of the SEM image as a function of exposure time. The acceleration voltage was 5~kV and the current 1.4~nA. The whole image sequence took about 20 s. The white scale bars correspond to a length of 5 $\mu m$. (d) Optical microscope image of bubbles exposed in the shape of an "hBN" label. The reddish dark shadow marks areas that were exposed by the electron beam. The white scale bar corresponds to a length of 20 $\mu m$. The red square indicates the region measured by AFM in (e). The AFM image shows that the wrinkle pattern vanishes on the bubbles due to strain relaxation, while it remains clearly visible elsewhere.}
	\label{fig:bubble}
\end{figure}

In this work, we grow hexagonal boron nitride by molecular vapor phase epitaxy (MOVPE) on two-inch sapphire substrates \cite{Kobayashi2008,Yang2018,Dabrowska2020, Ludwiczak2021}. For such hBN epilayers, we observed the formation of bubbles upon electron beam irradiation in vacuum in a scanning electron microscope (SEM). A typical example of bubble formation is shown in Figure~\ref{fig:bubble}~(c). The four images that were taken at different times (the whole sequence took about 20~s) show that first some tiny bubbles appear within the field of view and then gradually grow in diameter and height as the exposure continues. Eventually, bubbles may combine to form larger bubbles until all bubbles merge into a single large bubble that will rise roughly to the size of the exposed area (videos of the whole evolution of bubble formation are available in the supplementary information). An example of bubble formation can be seen in the optical microscope image, Figure~\ref{fig:bubble}~(d). Here, only the label "hBN" (marked by a dark shadow) was exposed to electron beam irradiation. One can clearly see that bubbles only form in or at the borders of the exposed areas and not on the whole sample. Interestingly, these bubbles remain stable even after the sample was removed from the high vacuum in the SEM and exposed to the ambient.  As can be seen in the AFM image in Figure~\ref{fig:bubble}~(b), hBN grown by MOVPE on sapphire shows many wrinkles. This behavior has already been reported \cite{Vuong2022,Dabrowska2020,Chugh2018} and can be ascribed to differences in the thermal expansion coefficients of hBN and sapphire, as shown schematically in Figure~\ref{fig:bubble}~(a). 

Since the hBN-layer is only weakly attached to the substrate, the strain induced during the cooling process leads to this typical wrinkle pattern. Bubble formation is one way to relax the strained wrinkle pattern locally, as can be seen in Figure~\ref{fig:bubble}~(e). Once the layer detaches, this strain redistribution stabilizes the bubble. Although it is clear that compressive strain favors bubble formation, the actual mechanism that leads to local delamination has yet to be identified. Important in this regard is that the bubbles form only in the irradiated areas, which means that they are a direct result of the electron-beam exposure. Because both the sapphire substrate and the hBN epilayer are insulating materials, electrostatic charging could be responsible. Indeed, charging effects are observed during SEM characterization and hinder high-resolution imaging. Another possible mechanism could be a chemical reaction involving gas evolution triggered by the electron beam. To shed more light on the actual mechanism, micro-Raman spectroscopy mapping was performed. 

\begin{figure}[h]
	\centering
		\includegraphics[width=1.00\textwidth]{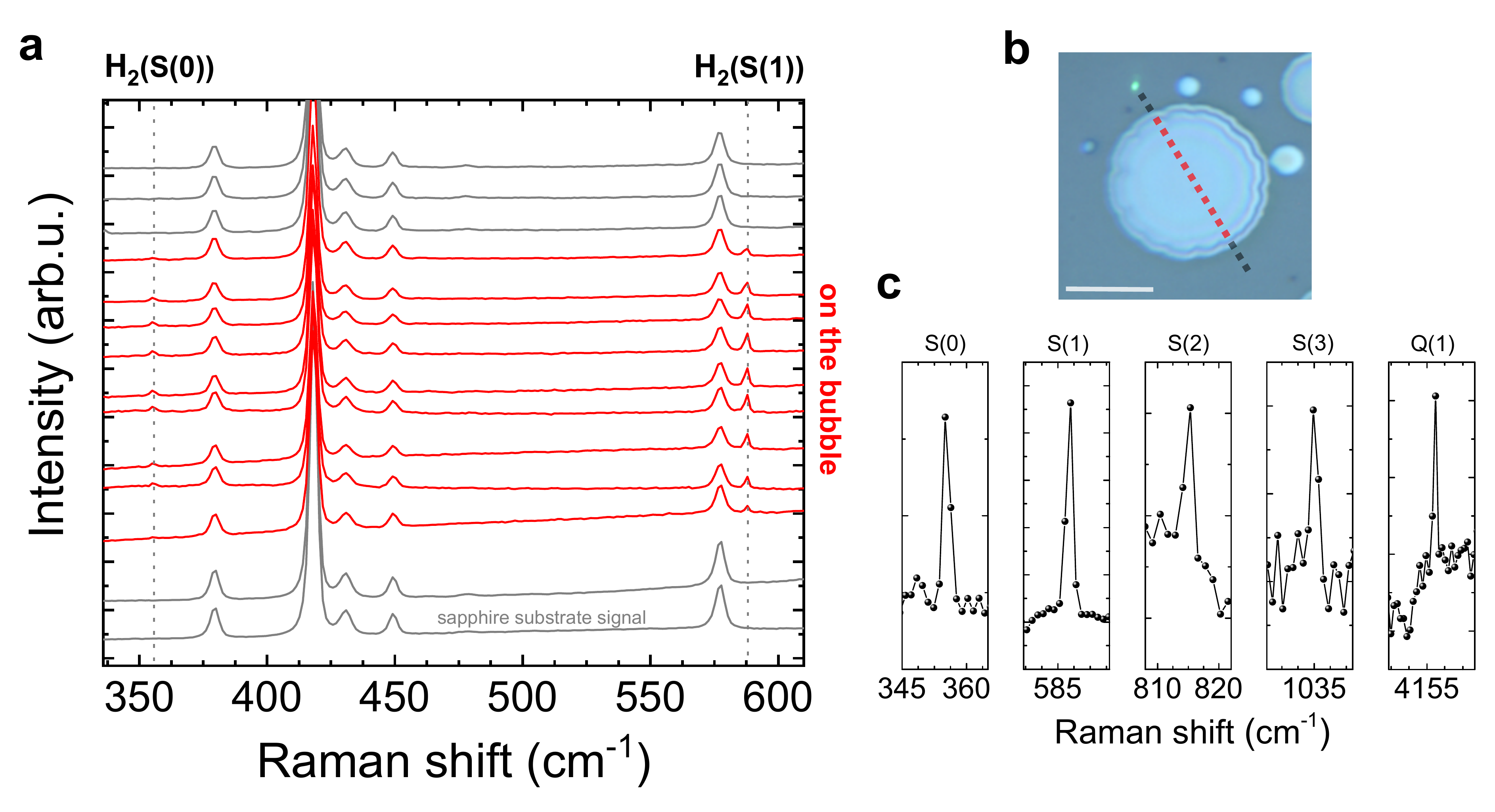}
	\caption{Evidence of molecular hydrogen in Raman spectroscopy. (a) Raman line scan across a hydrogen-filled bubble. It can be clearly seen that two lines are present only on the bubble. These lines correspond to the lines of molecular hydrogen S(0) at 354~$cm^{-1}$ and S(1) at 587~$cm^{-1}$. The Raman spectra are shifted vertically for clarity. (b) Optical microscope image of the studied bubble showing the laser spot (532~nm). The  dashed line indicates the direction of the line scan. (c) Additional rovibrational lines measured on the bubble not shown in the line scan. It was possible to identify the first four lines S(0)-S(3) of the (0-0) transitions and line Q(1) of the (1-0) transitions.}
	\label{fig:Raman}
\end{figure}

Figure~\ref{fig:Raman}~(a) shows the results of a line scan (step 2~$\mu$m) across a bubble as indicated in Figure~\ref{fig:Raman}~(b). The subsequent spectra are shifted vertically for the sake of clarity. Astonishingly, some very narrow Raman bands were only present when the laser was focused on the bubble. Two of these bands are marked in Figure~\ref{fig:Raman}~(a) by vertical dashed lines at 354~$cm^{-1}$ and 587~$cm^{-1}$. Figure~\ref{fig:Raman}~(c) presents all five peaks that could be clearly observed and that could be recorded only when the laser spot was on the bubble. The observed peaks can be ascribed to the S(0)-S(3) of the (0-0) transitions and the Q (1) line of the (1-0) transitions of molecular hydrogen \cite{Veirs1987,Campargue2012}. The observed presence of hydrogen inside the bubbles clearly points toward a chemical reaction triggered by the electron beam as a mechanism for bubble formation. 

\begin{figure}[h]
	\centering
		\includegraphics[width=0.9\textwidth]{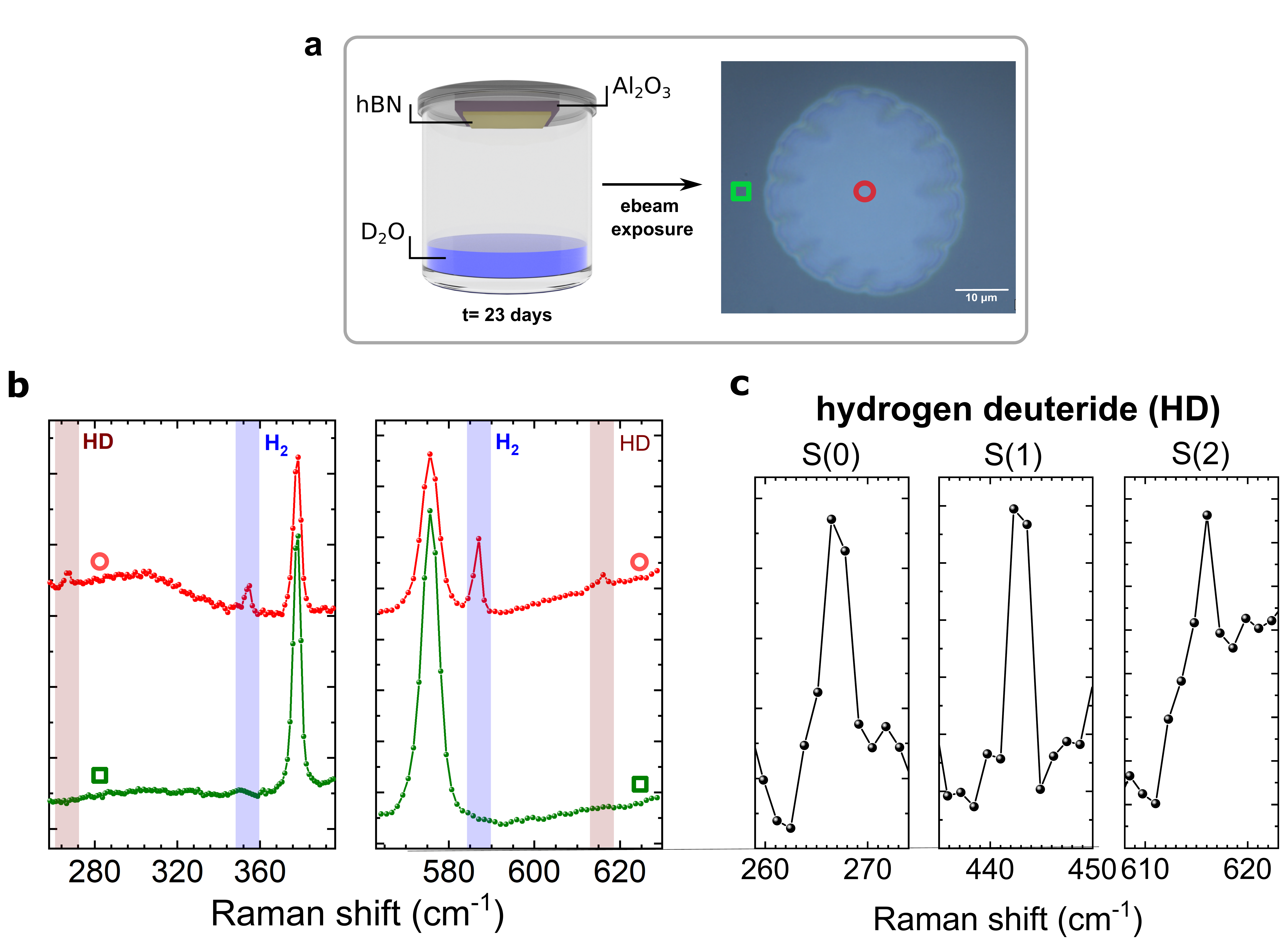}
	\caption{Origin of the hydrogen. (a) The sample was mounted upside down in a container filled with heavy water (D$_2$O) for 23 days. Afterwards the sample was directly mounted into the SEM and exposed by ebeam irradiation to form bubbles as shown in (a). (b) Raman spectra of two typical points on (next to) the bubble are shown as red (green) curves. The spectrum next to the bubble shows only Raman bands related to sapphire (the Raman band of hBN is at a higher energy, see  supplementary information). For the measurement on the bubble a signal related to molecular hydrogen, but also a signal related to hydrogen deuteride (HD) can be observed. (c) After subtracting the sapphire signal, three peaks (S(0), S(1) and S(2)) related to HD can be clearly observed. }
	\label{fig:HD}
\end{figure}

There are two possibilities to explain the source of hydrogen that leads to the formation of bubbles. First, the hydrogen gas is the result of the hydrogen accumulated in the layer during growth. Hydrogen is available in large amounts both in the precursor gases and in the carrier gas and can, for example, decorate defects or be stored in between layers. Second, the hydrogen source is introduced after growth, for example, by intercalation of water. To distinguish between these two possibilities, we placed a hBN/sapphire sample directly after growth inside a sealed container with heavy water. The sample was not in direct contact with the heavy water, but was mounted upside down at the top of the container. The container was kept closed for more than 20 days at room temperature to allow a possible intercalation of heavy water. After this period, the sample was removed and directly mounted inside the SEM for electron beam irradiation. Figure~\ref{fig:HD}~(a) shows a schematic drawing of this experiment. Similarly to the results shown in Figure~\ref{fig:bubble}, bubbles formed under irradiation. The samples were removed from the SEM and measured by micro-Raman spectroscopy. Raman bands associated with dimolecular hydrogen were identified, in agreement with Figure~\ref{fig:Raman}. However, in addition we observed novel bands at 267~$cm^{-1}$ and 616~$cm^{-1}$ that can be ascribed to the rovibrational lines S (0) and S (2) of hydrogen deuteride (HD), Figure~\ref{fig:HD}~(b) \cite{Stoicheff1957}. The S(1) line at 443~$cm^{-1}$ becomes also visible after subtracting the Raman signal from the sapphire, see Figure~\ref{fig:HD}~(c). The presence of deuterium clearly shows that intercalation occurs and that at least part of the hydrogen in the bubbles is connected to the water vapor introduced after growth, as was schematically shown in Figure~\ref{fig:bubble}~(a). This finding allows us to conclude that bubbles are formed as a result of the dissociation of water by electron irradiation (radiolysis). 
The first studies on the radiolysis of water date back to the beginning of the 20$^{th}$ century. The processes involved have been extensively studied and are of great importance, for example, for the safety and construction of nuclear plants \cite{LeCaeer2011}. More recent works show that the hydrogen generation yield by radiolysis of water can be greatly enhanced for certain wide-bandgap semiconductors  \cite{Petrik2001,Cecal2001,Southworth2018} and for water confined in nanostructures \cite{Rotureau2005,LeCaer2005}. Such enhancements would be needed to revisit the idea of hydrogen production via radiolysis of water using spent nuclear fuel \cite{Hart1976,Schneider2005}. More work is needed to determine the hydrogen generation yield of nanoconfined water in our hBN layers for different sources of ionizing radiation, but it is clear that the strained hBN/sapphire material system is exceptional because the formation of hBN bubbles allows for direct observation of the hydrogen gas production. 

\begin{figure}[h]
	\centering
		\includegraphics[width=1\textwidth]{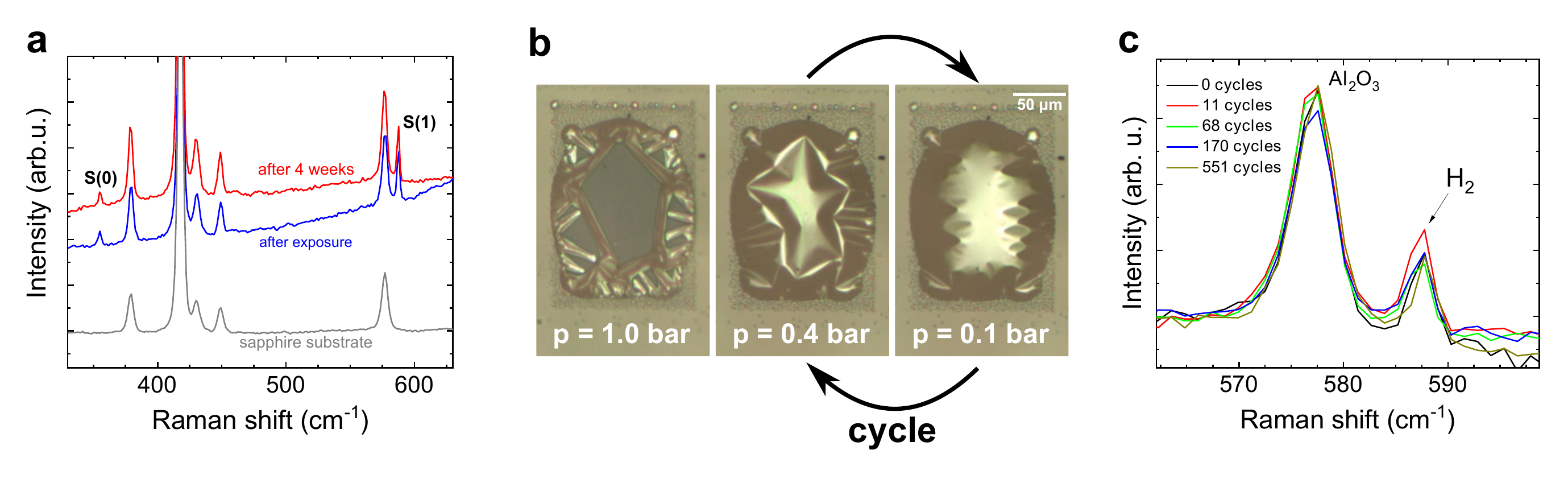}
	\caption{Evolution of the Raman signal of molecular hydrogen. (a) The graph presents two spectra taken at the same point of the bubble directly after electron beam exposure and after 4 weeks. The Raman lines did not decrease in intensity. This means that the hydrogen is still confined in the bubble after one month in ambient conditions. The Raman spectrum of a bare sapphire substrate is shown for comparison. (b) Optical microscope images of a large bubble for different ambient pressures. The bubble was inflated and deflated by automatically cycling the pressure between 100 and 400 mbar (see supplementary information for a video of a typical cycle). (c) After certain numbers of pressure cycles Raman measurements were performed. Molecular hydrogen was still present after 551 cycles. }
	\label{fig:stability}
\end{figure}

To explore storage capabilities, we monitored the Raman spectrum of H$_2$ over time. We found that H$_2$ can still be measured 4 weeks after exposure for a 40~nm thick layer (Figure~\ref{fig:stability}~(a)). After this time period, the bubble did not change visually under the microscope, but the H$_2$ signal disappeared. The H$_2$ generated after radiolysis, all other stable molecules (mostly H$_2$O$_2$)\cite{LeCaeer2011}, and the remaining interfacial water are confined to the bubble. Therefore, one could expect that back reactions would occur. In fact, such back reactions are among the limiting factors for H$_2$ production by radiolysis in pure water in closed systems \cite{LeCaeer2011}. The fact that we can observe H$_2$ over weeks points to an effective mechanism that prevents these back reactions. A similar suppression of back reactions was reported for the case of water confined in nanoporous materials \cite{Rotureau2005,LeCaer2005}, which indicates that the spatial confinement may also play an important role in our case. 

It is difficult to conclude whether the disappearance of the H$_2$ signal after weeks is due to slow back reactions or whether H$_2$ physically escapes the hBN layer. To be able to draw some conclusions about the suitability of our epilayers as a hydrogen barrier, we tested whether the structures remain leak-proof under extreme mechanical deformation. 
To this end, we made use of the fact that our bubbles are prepared under electron beam irradiation in vacuum. The shape of the bubbles shown in Figure~\ref{fig:bubble} remained almost unchanged after vacuum removal, which can be explained by the release of the strain in the wrinkles that stabilizes the layer when it detaches from the surface. However, bubbles will remain stable up to a certain diameter. Larger bubbles will collapse when they are removed from the vacuum of the SEM after illumination. An example of such a large bubble with a size of about 150~x~200~$\mu m$ is shown in Figure~\ref{fig:stability}~(b). Under atmospheric pressure, the bubble is compressed and folds down. However, when the sample is placed in a vacuum chamber and pumped down to 0.4~bar the bubble rises again, until it completely expands at a pressure of about 0.1~bar. To test the mechanical properties of hBN, we performed automated pump cycles between 0.1 and 0.4~bar leading to a compression and relaxation of the bubble. A video showing two cycles followed by a venting procedure to atmospheric pressure is shown in the supplementary information. Raman spectroscopy was used to measure the H$_2$ signal after a consecutive number of cycles. Figure~\ref{fig:stability}~(c) shows that the signal of H$_2$ remained unchanged,  within the accuracy of the Raman measurement, even after the maximum number of 551 cycles. Therefore, we can conclude that the mechanical strength of large-area epitaxial hBN by MOVPE holds great promise for application as a hydrogen barrier, for example, for future H$_2$ lightweight storage tanks.

In summary, we show that epitaxial hBN grown by MOVPE on two-inch sapphire substrates is a prospective material for hydrogen generation and storage. hBN bubbles are formed upon electron beam irradiation, and Raman spectroscopy shows the presence of molecular hydrogen. Experiments with heavy water provide evidence that hydrogen generation is triggered by the radiolysis of interfacial water. The hydrogen produced in these bubbles is detectable for weeks, despite possible back reactions. The bubbles remain leakproof even under intense mechanical deformation, which highlights the flexibility and mechanical strength of the epitaxial hBN grown by MOVPE.  More work is needed to study whether the splitting of interfacial water can also be achieved by other types of radiation (e.g. UV light), which could further expand the field of application, but the presented results already indicate that epitaxial hBN has the potential to be applied in future innovative hydrogen generation and storage schemes.
\section{Methods}

\subsection*{Boron nitride growth}
The boron nitride layers were grown on two-inch sapphire substrates (c-plane) by Metalorganic Vapour-Phase Epitaxy (MOVPE) using an Aixtron CCS 3x2-inch reactor equipped with an ARGUS Thermal Mapping System to directly control the substrate temperature. Growth precursors are triethylborane (TEB) and ammonia. Hydrogen was used as carrier gas. The layers were grown by pulsed growth and two-stage epitaxy \cite{Dabrowska2020} at a temperature typically between 1260 -1300 $^{\circ}$C. 

\subsection*{Characterization}
Micro-Raman spectra were recorded using a Renishaw inVia Raman setup equipped with a 532~nm continuous wave laser excitation source and a 100x objective. The spot size was about 1~$\mu$m with a laser power density of 3~$\cdot 10^{5}$~$W/cm^{2}$. The system has an automated xyz stage with a resolution of 100~nm.  SEM imaging and ebeam exposure were performed on a FEI Helios 600 Dual Beam system, equipped with a Raith Elphy electron lithography setup. AFM images were acquired with a Bruker Dimension Icon microscope / Nanoscope VI controller in a PeakForce mode. Bruker ScanAsyst Air probes were used for imaging.

\section{Data availability}
The data that support the findings of this study are available from the corresponding authors upon reasonable request.

\bibliography{hydrogen}

\begin{acknowledgement}
This work was supported by the National Science Centre,
Poland, grants 2019/ 33/B/ST5/02766 and
2020/39/D/ST7/02811.

\end{acknowledgement}

\begin{suppinfo}
The Supporting Information contains additional Raman and X-ray diffraction measurements,  two videos showing the bubble formation in-situ in a scanning electron microscope and a video showing the pumping cycles (bubble deformation) recorded in an optical microscope. 

\end{suppinfo}


\end{document}